# Mesoporous Thin-Films for Acoustic Devices in the Gigahertz Range

N. Lopez-Abdala[1], M. Esmann[2], M. C. Fuertes[3], P. C. Angelomé[3], O. Ortiz[2], A. Bruchhausen[4], H. Pastoriza[4], B. Perrin[5], G.J.A.A. Soler-Illia[1, †], N. D. Lanzillotti-Kimura[2, ‡]

[1]Instituto de NanoSistemas – Universidad Nacional de San Martín-CONICET,
Buenos Aires, Argentina
[2] Université Paris-Saclay, CNRS, Centre de Nanosciences et de Nanotechnologies,
91120 Palaiseau, France
[3] Gerencia Química & Instituto de Nanociencia y Nanotecnología, Centro Atómico Constituyentes,
CNEA-CONICET, Buenos Aires, Argentina
[4]Centro Atómico Bariloche & Instituto de Nanociencia y Nanotecnología, CNEA-CONICET,
Rio Negro, Argentina
[5]Sorbonne Université, CNRS, Institut des NanoSciences de Paris, INSP, F-75005 Paris, France

† email: gsoler-illia@unsam.edu.ar
‡email: daniel.kimura@c2n.upsaclay.fr

*The coherent manipulation of acoustic waves on the nanoscale usually requires multilayers with thicknesses and interface roughness defined down to the atomic monolayer. This results in expensive devices with predetermined functionality. Nanoscale mesoporous materials present high surface-to-volume ratio and tailorable mesopores, which allow the incorporation of chemical functionalization to nanoacoustics. However, the presence of pores with sizes comparable to the acoustic wavelength is intuitively perceived as a major roadblock in nanoacoustics. Here we present multilayered nanoacoustic resonators based on mesoporous $SiO_2$ thin-films showing acoustic resonances in the 5-100 GHz range. We characterize the acoustic response of the system using coherent phonon generation experiments. Despite resonance wavelengths comparable to the pore size, we observe for the first time unexpectedly well-defined acoustic resonances with Q-factors around 10. Our results open the path to a promising platform for nanoacoustic sensing and reconfigurable acoustic nanodevices based on soft, inexpensive fabrication methods.*

Keywords: mesoporous thin-films, picosecond ultrasonics, nanomechanics, nanophononics, acoustic-phonons, nanocavities

## INTRODUCTION

Nanophononics and nanomechanics deal with vibrations in solids where the typical frequencies are in the range spanning from few GHz up to THz.[1]–[12] These frequencies have associated wavelengths in the 1-100 nm range, making them interesting for high-resolution acoustic imaging and sensing applications. Ultrasound imaging in the kHz-MHz has revolutionized medical applications; a similar impact can be predicted for nanoacoustic waves for non-destructive testing at the nanoscale[13]–[15]. The design and engineering of acoustic interference devices have been traditionally based on almost atomically-flat interfaces presenting roughness only well below the phonon wavelength. The individual nanometric layer thicknesses are usually very well defined.[5], [16] Naturally adapted to these requirements, the vast majority of the works on nanophononics are based on samples grown by molecular-beam epitaxy (semiconductor, oxides)[3], [16]–[21], and processed using electron beam lithography (plasmonic and micropillar structures)[6], [9], [22]–[25]. This results in expensive devices with predetermined functionality.

Colloidal crystals or polymer and mesoporous multilayers seem equally promising for new developments in phononic crystals or metamaterials.[4], [26] For example, hypersonic one-dimensional crystals were made from alternating layers of poly(methyl methacrylate) and porous silica.[27] as well as multilayered porous silicon[28] More recently, a new class of tunable hypersonic phononic crystals has been developed, based on economically viable polymer-tethered colloids that present a tunable hybridization gap.[29] These systems opened the path towards the integration of soft matter-derived materials into nanoacoustic devices.

The fabrication of mesoporous thin films is an efficient, cheap, and reproducible technique. Mesoporous materials present great versatility for optical sensing applications[30], [31] and feature the capability of hosting fluids and particles in their pores. For instance, porous networks from a wide variety of metal-oxides in crystalline and amorphous states can be designed, offering tailored electronic and optical properties, like ultra-low dielectric constants and photocatalytic activity.[32]–[35]

All these characteristics could be useful in the nanoacoustic domain to develop tunable nanoacoustic devices and sensing applications.[36] Despite its potential, the development of high-frequency acoustic devices based on mesoporous materials has received little attention from the scientific community. Mesoporous materials are perceived as diffuse media since their pores have sizes comparable to the acoustic wavelength. Intuitively, this should prevent the fabrication of interference-based nanoacoustic devices.

Here we demonstrate multilayered nanoacoustic resonators based on mesoporous $SiO_2$ thin-films showing acoustic resonances in the 5-100 GHz range. We characterize the acoustic response of the system using coherent phonon generation experiments. Despite resonance wavelengths comparable to the pore size, we observe unexpectedly well-defined acoustic resonances with Q-factors around 10.

## SAMPLE FABRICATION AND CHARACTERIZATION

Three samples were fabricated (named A, B, and C), each one consisting of three layers. A mesoporous $SiO_2$ thin film was first deposited on top of the substrate, and a capping layer of dense $SiO_2$ was deposited after a stabilization process. Subsequently, a thin film of Ni (~32 nm) was deposited by evaporation to act as a transducer in the nanoacoustic characterization experiments. The dense silica layer prevents the diffusion of the metal into the pores. We also fabricated a control sample without the mesoporous silica layer. The thickness and mesoporosity parameters of the layers, along with the withdrawal speeds, are presented in Table 1. The thickness and mesoporosity were deduced from ellipsometry measurements (SOPRA GES5E ellipsometer). All the sample preparation steps and characterization techniques are described in detail in the SI.

|   | Mesoporous layer | | | Dense layer | |
| --- | --- | --- | --- | --- | --- |
|   | Thickness (nm) | Porosity (%) | Withdrawal speed (mm/s) | Thickness (nm) | Withdrawal speed (mm/s) |
| A | 109 | 42 | 2 | 30 | 0.2 |
| B | 137 | 40 | 3 | 25 | 0.2 |
| C | 161 | 39 | 4 | 35 | 0.2 |
| Control | - | - | - | 50 | 1 |

Table 1: Samples properties and dip-coating withdrawal speeds. Errors are on the order of 5% for all the reported values.

Figure 1a shows a typical transmission electron microscopy (TEM) micrograph of mesoporous silica. In this image, the oxide walls are seen as darker areas while the lighter areas correspond to the pores. The observed pore array is compatible with the typical structure of F127-templated mesoporous oxides: a body-centered cubic array of pores, that in some cases, is slightly distorted, giving rise to regions in which the order is less evident. Figure 1b shows a typical scanning electron microscopy (SEM) image of the layer structure in sample A. A schematic sample cross-section is included as a guide to the eye. The image was taken before depositing the metallic transducer. The substrate, mesoporous, and dense layers can be clearly identified. By analyzing large cross sections, we observe that the thickness of the mesoporous layer is more uniform and better defined than the dense layer (see Supplementary Information S2).

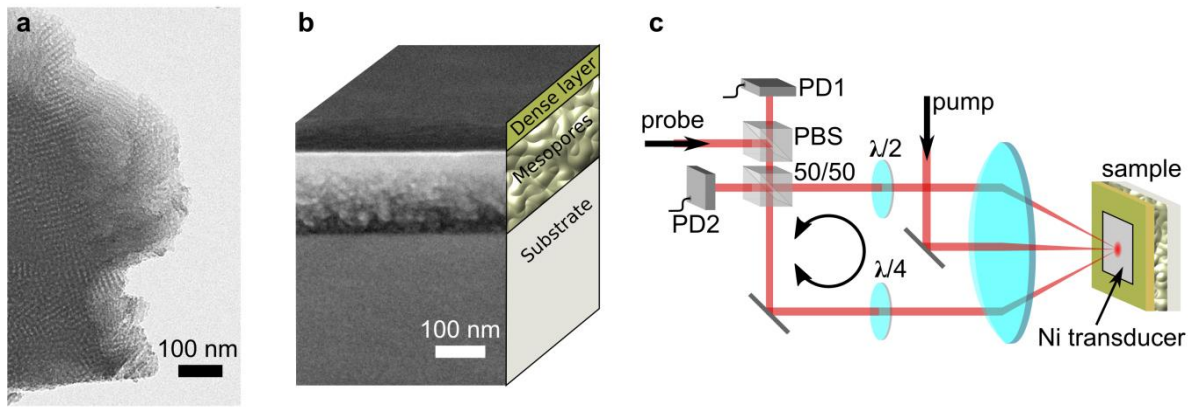

Figure 1: **a** TEM image of $SiO_2$ mesoporous film showing the mesopore array. **b** SEM micrograph of sample A showing the substrate, the mesoporous, and the dense layers. On the right, a schematic of the sample is presented as a guide to the eye. **c** Interferometric pump-probe setup. A strong pump pulse excites acoustic phonons through partial absorption in a Ni transducer layer. A probe pulse detects the time-resolved surface displacement using a Sagnac interferometer.[37], [38]

**NANOACOUSTIC CHARACTERIZATION**

To acoustically characterize the samples, we performed coherent-acoustic-phonon generation experiments[39], [40] using the setup sketched in Figure 1c. This pump-probe technique allows the study of ultra-high frequency acoustic phonons traveling through a material. The time-resolved analysis of the optical reflectivity of the sample perturbed by the presence of coherent phonons enables the study of phonon dynamics on the picosecond time scale. In this technique, a short, intense laser pulse (pump, λ=758nm, 200fs pulse duration, 80 MHz repetition rate, 275mW power) is absorbed by a transducer (Ni thin-film in our case), generating a broadband wavepacket of coherent acoustic phonons mainly by the thermoelastic effect. Usually, optically transparent materials such as $SiO_2$ are very inefficient optoacoustic transducers. The transducer material and thickness are crucial engineering parameters of the device. To permit the injection of phonons into the mesoporous layer, a material with an acoustic impedance sufficiently similar to that of $SiO_2$ is needed. Nickel meets this requirement reasonably well. Gold would represent a feasible alternative considering its acoustic impedance.

The propagation of phonons into the sample is governed by its frequency-dependent acoustic response. Upon propagation, the phonons modulate both the optical properties of the materials constituting the studied structure and the position of the interfaces. In our particular case, we implemented a Sagnac interferometric detection scheme that allowed us to recover mainly the contributions of the surface displacement to the change of optical reflectivity.[37], [38]

A delayed, less intense pulse (probe) measures the instantaneous optical reflectivity as a function of the time delay between both pulses. Any resonance in the acoustic response of the sample thus induces a periodic modulation in the detected reflectivity time-trace. By performing a Fourier transform of this signal, the frequencies of the coherent acoustic phonons present in the sample can be determined. In the case of the interferometric detection, two counterpropagating pulses probe the sample and are recombined in two photodetectors (PD1 and PD2). To recover the surface displacement signal, the two cross-polarized probes reach the sample at slightly different times. The half-wave plate and the quarter-waveplate ensure the cross-polarization of the beams and maximize the sensitivity of the interferometer.[37], [38]

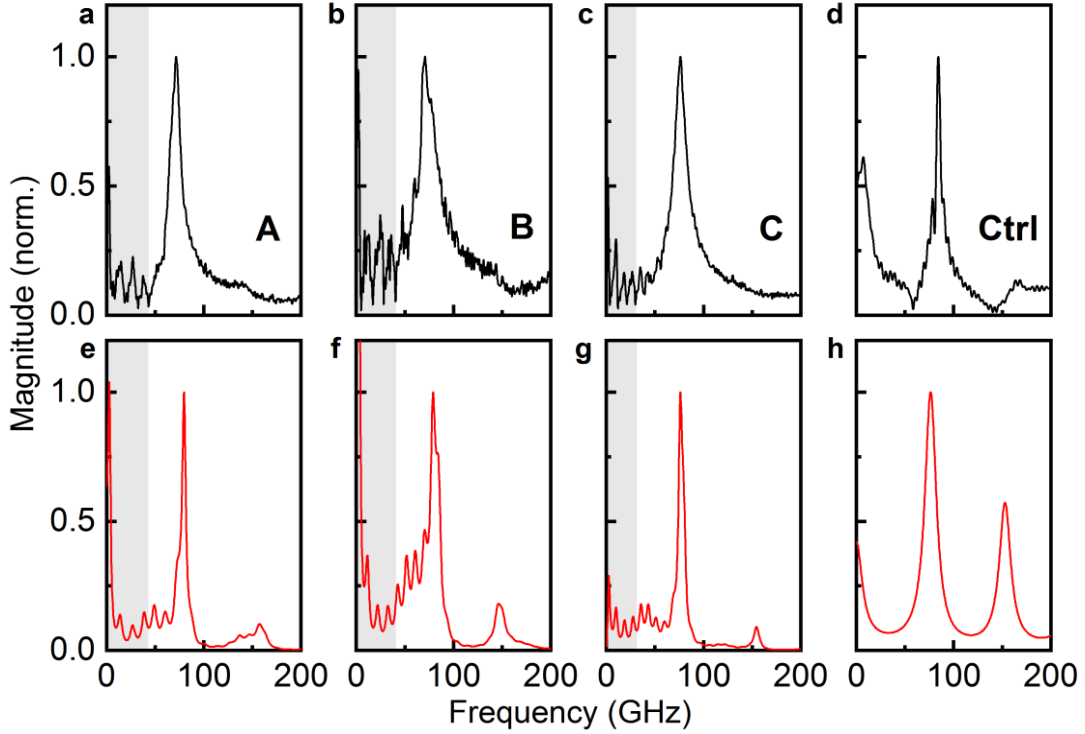

Figure 2: Coherent phonon generation spectra in mesoporous acoustic resonators. Panels a, b, c, and d correspond respectively to the samples A, B, C, and the control sample (Ctrl) without a mesoporous layer, respectively. The top row presents the measured spectra obtained from Fourier transforms of transient reflectivity traces. The bottom panels show the corresponding simulated spectra. The peaks in the low-frequency range correspond to the multiple resonances in the mesoporous thin film. The grey regions are guides to the eye, indicating the three first harmonics supported by the mesoporous layer. The intense peak appearing around 80 GHz in all spectra corresponds to a resonance in the Ni transducer layer.

In Fig. 2, we show the results of the measured (top) and simulated (bottom) spectra of the time-resolved reflectivities in the studied mesoporous samples. In all spectra, an intense peak appears around 80 GHz. As we will show later, this peak is associated with a joint acoustic resonance of the metal transducer and the dense cap layers. On the lower frequency side of this intense peak, we observe a series of clearly defined oscillations. They correspond to acoustic phonons confined in the mesoporous layer. The grey regions are guides to the eye, indicating the three first harmonic supported by the mesoporous film. For an isolated film, the mode spacing is purely determined by the mesoporous film thickness $d_M$ and the speed of sound in the mesoporous material $v_M$ via $\Delta f_M = v_M/2d_M$. As a rule of thumb, a thicker film thus results in more densely spaced resonances, i.e., a larger number of visible peaks to the left of the main resonance around 80 GHz.

For sample C (Fig. 2c), up to six harmonics of these modes can be distinguished in the experimental spectrum, at frequencies: 9.9, 18.2, 26.7, 34.9, 42.9, and 52.8 GHz. For samples A and B, shown in Fig. 2a and 2b, respectively, the oscillations are still observable, yet less defined than in Fig. 2c. In contrast to samples A-C, the experimental spectrum of the control sample (Fig. 2d) does not show any of these oscillations featuring only the metal layer resonance around 80 GHz.

To account for the experimental results, we performed simulations of the pump-probe experiments using a transfer matrix method.[5] The resulting simulated spectra of surface displacement (i.e., the same quantity measured experimentally) are shown in Fig. 2e-h. We see a good agreement between experiments and simulation in number, position, and relative height of the peaks.

The transient optical reflectivities are governed by the optical and acoustic response of the sample. That is, the relevant parameters for the simulation are the geometric layer thicknesses in Table 2, together with the corresponding refractive indices, speeds of sound, and mass densities. For the substrate, the dense $SiO_2$, and the Ni layer, these parameters are well known and summarized in table S1 of the SI. For the mesoporous layer, we derive the optical and elastic properties as weighted averages of $SiO_2$ (dense matrix) and air (pores) with the porosity determining the relative composition. We find the closest agreement between experiment and simulation for a refractive index of $n_{meso} = 1.32$ (at λ=700nm-1000nm) and speed of sound of $v_{meso} = 3157 \ m/s$. We furthermore consider the substantial acoustic losses in the mesoporous material with an effective phonon decay length of $\delta_{meso} = 90 \ nm$. Surprisingly, we find that the best match between experiment and simulation is obtained for a mesoporous mass density of $\rho_{meso} = 2.2 \ g/cm^3$, i.e., the same density as the glass substrate and the dense $SiO_2$. That means the effective acoustic impedance mismatch (determined by $Z = \rho \cdot v$) between the mesoporous material and the adjacent layers is much lower than anticipated from a simple weighted impedance average. We hypothesize that this effect may be a consequence of the *internal* impedance mismatch between the host material ($SiO_2$) and the nanopores (air). Essentially, sound in the GHz-THz regime does not propagate in air, hence percolating only through the host material. This hypothesis still holds true if considering that the pores are partially filled with water since the experiments were performed under ambient conditions.

As a consequence, the much lower impedance mismatch between the host matrix and glass/$SiO_2$ determines the acoustic transmission from the mesoporous material into the adjacent layers. To test this hypothesis, a systematic study of the variation of the mesoporous layer thickness while keeping all other parameters constant should be performed. This study, however, goes beyond the scope of the present manuscript.

Figures 2d and 2h show the control sample experimental and simulated spectra, respectively. The lack of low-frequency modes is a contrasting result in comparison with the samples embedding a mesoporous layer (Fig. a-c and Fig. e-g). Additionally, we can see a slight red-shift of the simulated spectrum with respect to the experimental one. Since the metal layer was deposited during the same process for the four samples, we conclude that the difference is mainly due to the non-uniformity of the dense $SiO_2$ layer, and possible penetration of the metal into the oxide layer. This hypothesis is supported by SEM images, which show in general an increased roughness in the surface of the dense layer compared to the mesoporous film (Supplementary Information S2).

To further confirm that the nature of these peaks is related to the mesoporous layer, we simulated the acoustic mode profiles corresponding to the three first measured peaks in sample C. Figure 3 shows the profiles, from bottom to top, corresponding to 10, 19, 28 GHz and for the central peak at 77 GHz. The curves have been vertically shifted for clarity. We see that the modes at 10, 19, and 28 GHz indeed correspond to the first three harmonics of the confined standing wave modes in the mesoporous layer, showing 1, 2, and 3 maxima, respectively. These three modes are indicated with a grey guide to the eye in Fig. 2. For higher harmonic orders, the distribution of energy among all the layers prevents a precise localization in the mesoporous layer. For the mode at 77 GHz, there are 8 maxima within the mesoporous layer. However, most remarkably, this mode features a standing wave within the metal layer and the dense $SiO_2$ layer below, which together act as a resonator with an acoustic thickness of one wavelength.

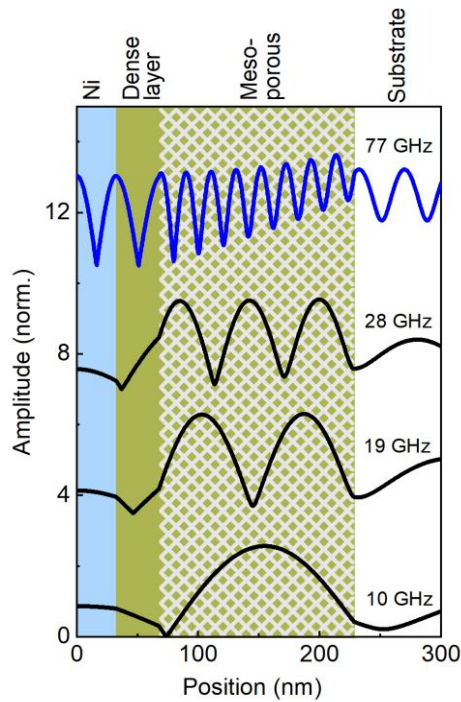

Figure 3: Numerical simulation of the mode-spatial profiles corresponding to the sample C. The curves from bottom to top correspond to 10, 19, 28, and 77 GHz, respectively. The curves have been vertically shifted for clarity. The three lowest-frequency modes show an evident standing wave character in the mesoporous layer.

**CONCLUSION AND PERSPECTIVES**

For the past 20 years, there has been a growing interest in mesoporous materials due to their high surface area -of the order of 1000 $m^2/g$- the precise control in the pore size and shape, pore arrangement, and tunable surface functionality.[41], [42] In particular, mesoporous thin films obtained from soft chemistry methods present increasing interest as highly accessible materials with tunable structural, chemical electronic and optical properties that open the path to applications in fields such as sensing, environment, energy, and biomedicine.[34], [35] The optical properties of mesoporous films are easily tunable through physical and chemical methods and are subject to change when exposed to the environment, which opens the field of responsive thin-films acting as optical waveguides.[43] Besides, complex architectures such as mesoporous multilayers can be processed as 1D photonic crystals or optical resonators with a selective optical response to vapors with different chemical properties or molecular size.[30], [44], [45] This application exemplifies that molecular, supramolecular, or photonic information can be encoded into the structure of a multilayer at several length scales[6], or indeed combined with the plasmonic response.[46] The extension of all these concepts into the nanoacoustic domain represents promising perspectives of this work.

In this work, we have shown that mesoporous materials with pore sizes of the order of 10 nm can support stationary modes with frequencies in the 5-100 GHz range with typical Q-factors of 10. The chosen device geometry is the simplest one to demonstrate the feasibility of the use of mesoporous materials in nanoacoustic applications.

We anticipate that the combination of a wall structure and pore size control with the possibility of including molecular or nanosized species within the mesopores opens the gate for highly integrated nanophononic devices responsive to external stimuli and to exploit confinement effects in mesopores.[47] This integration of physical and chemical phenomena at several length scales can be projected to optoelectronics or optomechanical applications of responsive matter.[48]

**SUPPLEMENTARY INFORMATION**

**S1) SIMULATION PARAMETERS**

The simulations were performed using an implementation of the transfer matrix method[16], [49], both for the electrical field and the acoustic displacements. The following Table summarizes the material parameters used simulations of pump-probe spectra shown in Figure 2 (bottom row) of the main text. These parameters were used in conjunction with the nominal layer thicknesses summarized in Table 2. The simulation was performed following the approach detailed in Ref. [16], [39], [49], [50]. We furthermore assumed that optical absorption and coherent phonon generation are entirely limited to the Nickel transducer layer.

|  | Refractive index | Speed of sound (m/s) | Mass density (g/cm^3) | Acoustic decay length (nm) |
|---|---|---|---|---|
| Air | 1.00028 | 343 | $1.275 \; 10^{-3}$ | - |
| Nickel | 2.218+4.8925i | 4970 | 8.908 | - |
| Dense $SiO_2$ | 1.5375 | 5750 | 2.2 | - |
| Mesoporous | 1.32261 | 3156.7 | 2.2 | 90 |
| Glass | 1.5375 | 5750 | 2.2 | - |

Table S1: Optical and acoustic material parameters used for the simulation in Figure 2 of the main text.

**S2) SAMPLE SURFACE ROUGHNESS**

To characterize the surface and interface roughness of the samples presented in the main text, we recorded a large scale SEM image as displayed in Fig. S1. This image was taken before the deposition of the Ni transducer layers. We observe an increased roughness at the top surface of the dense layer in comparison to the surface facing the mesoporous film below. We hypothesize that this roughness is partly responsible for the observed red-shift of the principal peak around 80GHz in the control sample (Fig. 2d of the main text) with respect to the calculated counterpart based on nominal thicknesses (Fig. 2h).

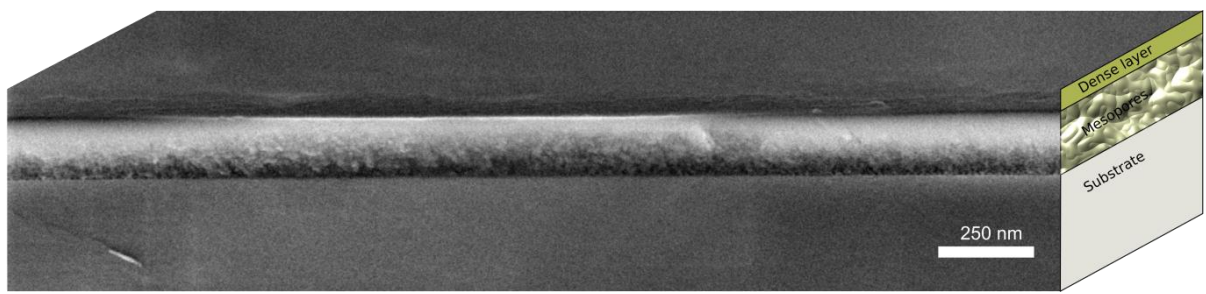

Figure S 1: Large scale scanning electron micrograph of sample A showing the substrate, the mesoporous, and the dense layers. On the right, a sketch of the sample structure is included as a guide to the eye. The interface between the mesoporous layer and the dense SiO2 layer is much smoother than the top surface of the dense layer. The SEM image was taken before Ni deposition.

**S3) METHODS**

Sample fabrication:

Mesoporous thin silica films were produced using the Evaporation-Induced Self-Assembly approach.[34] Silica dense and mesoporous layers were deposited onto glass substrates by dip-coating from absolute ethanol-water solutions. Tetraethyl orthosilicate (Sigma-Aldrich) was used as the oxide precursor and F127 block copolymer $(EO)_{106}(PO)_{70}(EO)_{106}$ (Sigma-Aldrich) as a mesopores template. A prehydrolysis step is needed to optimize the chemical reactivity of the silica precursor. It consists of a 1-hour-reflux of an initial solution of TEOS:EtOH:$H_2O$ with a 1:5:1 molar ratio. After this step, the final solutions for dense and mesoporous films are prepared by adding solvent, water and the template (when necessary) The final mixture, with a molar relation of TEOS:$H_2O$:HCl:EtOH:F127 of $1:10:0.25:40:5 \times 10^{-3}$ was aged at room temperature for 48 h under agitation. A solution with the same composition but no added F127 was also prepared in the same conditions, to deposit dense (i.e., non-mesoporous) silica as an overlayer, or as a control film.

After dip-coating, the post-treatment for mesoporous layers involves an aging and thermal stabilization process, that consisted of three 30-min steps, first in a chamber at room temperature and 50 % relative humidity, and then thermal treatment at 60 and 130 °C, followed by a consolidation step of 2 h at 200 °C. This stabilization treatment is essential to obtain ordered mesoporosity and to consolidate the mesoporous film structure to keep its integrity when doing the dip-coating of the second layer.[46]

The thermal treatment for the dense layers consisted on a single step of 200 °C for 2 h, and for the samples with mesoporous silica, a final calcination step of 2 h at 350 °C with a 1 °C/min heating ramp is carried out, to eliminate the pores template.

The nickel cover layers were deposited by thermal evaporation using a home-made vacuum thermal evaporation system. The base pressure during evaporation was $8 \times 10^{-7}$ Torr. Nickel was evaporated by placing a piece of 99.95% Ni rod (Goodfellow) on a alumina covered Tungsten boat. Film thickness was measured during the deposition by monitoring the frequency change of a resonant quartz crystal.

TEM and SEM images:

The TEM images were obtained in a JEOL JEM2010 FEG TEM operating at 200 kV. Samples were prepared by scratching off the films from the substrate and depositing them on FORMVAR and carbon-coated copper grids.

The SEM images were obtained in an FE-SEM Zeiss supra 40 equipped with a field emission gun. Samples were cleaved prior to imaging of the resulting vertical cross-section.

**Acknowledgments**

NDLK, ME, NLA, and OO acknowledge funding by the European Research Council Starting Grant No. 715939, Nanophennec. M.E. acknowledges funding by the Deutsche Forschungsgemeinschaft (DFG, German Research Foundation) Project 401390650. GJAASI acknowledges ANPCyT for projects PICT 2015-3526 and 2017-4651. MCF acknowledges ANPCyT for projects PICT 2015-0351 and 2017-1133. NDLK and GJAASI acknowledge funding from CNRS through the project IEA RANas.